# Group Representational Clues to
# A Theory Underlying Quantum Mechanics

Casey Blood
Sarasota, FL
CaseyBlood@gmail.com

## Abstract
The current form of quantum mechanics is very successful and is almost certainly "correct." It is remarkable, however, that the entire structure—from the mass, spin and charge labels on particle states to antisymmetry to broken internal symmetries to gauge transformations to the equations of motion—is built upon concepts from group representation theory. That is, the theory is constructed *exactly as if* it were a representational form of an underlying theory. Our proposed form for the underlying theory is that it is based on a linear equation, $O\text{F}(V)=0$. F is a function of, some set of "independent," currently unknown variables, $V$, with $O$ being a linear, partial differential operator in those variables. The operator is assumed to be invariant under a group of transformations of the $V$'s, with the invariance group being homomorphic to the direct product of the inhomogeneous Lorentz group and the internal symmetry group. In such a theory, a state vector, denoted by a ket with group-theoretic labels, would represent a function of the independent variables. In addition to explaining the group representational structure of quantum mechanics, an underlying theory offers insight into gauge theory.

## Prologue

The question we are concerned with here is whether quantum mechanics is the final theory of the physical universe. Our answer carries with it a re-evaluation of the nature of physical reality. In book VII of the Republic, Plato paints an image of the state of humankind. We are like prisoners in a cave, chained so we cannot even move our heads. We face a wall at the back of the cave on which are thrown shadows of the events in the real world—people walking, trees blowing in the wind and so on. These shadows are a *representation*—a re-presentation—of the real world, but they are not the real world.

What we will infer here is that we are in the same state as the prisoners in the cave. Quantum mechanics will be shown to imply that in the "real" or most basic physical realm, there is neither space nor time nor matter. Instead these concepts are like the shadows on the wall of the cave, a representation of an underlying reality.

How could we possibly infer this from quantum mechanics, which seems to be such a complete and successful theory as it is? There is a branch of mathematics called, appropriately enough, representation theory. In its original form, a mathematical equation will have such and such a form. But one can construct another equation which is a *representation* of the original equation. It has the same structure as the original equation in many respects, but not in all. There will be certain clues in the structure of the equation that it is indeed a representation of the original and not the original itself. We show here that quantum mechanics has exactly the characteristics one would expect if it were a representation of a deeper theory. Space, time and matter do not appear in the original statement of this deeper theory; instead they are characteristics of the *solutions* of the equation. It is the solutions we perceive, and we then construct our mental world around the characteristics of those solutions.

## 1. Introduction.

Is quantum mechanics as it now stands the final theory of the physical universe? Its highly unified mathematical structure and its successes in elementary particles, atomic and nuclear structure, and solid state physics certainly seem to imply it is on the right track. On the other hand, the current theory of quantum mechanics is built *entirely* upon concepts from representation theory. The particle-like properties of mass, energy, momentum, and spin—associated with the inhomogeneous Lorentz group *ISL*(2) [1]—and charge—associated with the internal symmetry group *G*—are group representation labels; antisymmetry is a concept from representations of the permutation group; unbroken and broken internal group symmetries and their representations play a prominent role in elementary particle physics; and the free-particle Dirac and Proca (spin 1) equations follow from group representation theory.

This ubiquitous representational structure would seem to imply with near certainty that there is a more fundamental form of quantum mechanics, with the current form being a representation of the underlying theory. To conform to the basic principles of quantum mechanics, the underlying, pre-representational theory must be linear. And it must be invariant under a set of transformations that is homomorphic to the direct product of *ISL*(2), the internal symmetry group *G*, and a permutation group. The general structure of such a theory is given in section 2, with a specific example given in appendix A.

Representation theory is briefly reviewed in section 3. The basis used to change from the pre-representational to the current representational form of the theory is discussed in section 4. The vacuum is presumed to be made up of extended "molecules" that are invariant under global but not local internal symmetry operations. Gauge transformations [2] are then used to construct vector boson states from modification of these molecules. This gives one an understanding of why gauge transformations generate vector boson fields. It explains why the fields obey symmetric statistics, why the gauge-generated field is proportional to the derivative of the gauge transformation, why there is one vector boson for each generator of the invariance group, and why fields have the same transformation properties as the generators of the gauge group. Finally, the interpretation of the theory is discussed in section 5.

## 2. General Form of the Underlying Theory

We are looking for an underlying theory that has a *representational form* identical to the current theory of quantum mechanics. To accommodate the structure of the current theory, the underlying *pre-representational* theory must be linear, it must be Lorentz invariant, it must have an internal symmetry group, and it must naturally allow for antisymmetry. We conjecture that the underlying theory has the following form:

**(1)** All physically relevant states correspond to solutions of a linear equation

(1)
$$O\Psi(\eta^1, \eta^2, ...) = 0$$
$$O = \sum_{i=1}^{N_0} O^{(1)}(\eta^i) + \sum_{i \neq i'} V(\eta^i, \eta^{i'}).$$
$$V(\eta^{i'}, \eta^i) = V(\eta^i, \eta^{i'})$$

$O$ is a linear operator, perhaps a partial differential operator, in some currently unknown set of underlying independent variables, $\eta^i, i = 1, 2, ..., N_0$, with the interaction term, $V(\eta^i, \eta^{i'})$, being a symmetric function of (and/or operator on) the two sets of variables $\eta^i$ and $\eta^{i'}$.

The underlying Eq. (1) is pre-representational in the sense that $\Psi$ does not correspond to a particular representation. The different solutions $\Psi$ correspond to all possible physical states. $\Psi$ may correspond to just one particle or to many particles, including both fermions and bosons. For a particular $\Psi$, its representation in some particle-like basis will correspond to the total state vector of that state. It is the *representation* of $\Psi$, rather than $\Psi$ itself, that we deal with in current quantum mechanics.

We do not know the form of the operator $O$, or even the nature of the underlying independent variables. However, we give an illustrative example [3] (see also [4]) in appendix A in which each $\eta^i$ corresponds to a set of $4n'$ *complex* variables, with the operator $O^{(1)}$ being a second order differential operator. Different solutions correspond to different masses and spins.

**(2)** There is a group of transformations of each set of variables $\eta^i$ that is homomorphic to the direct product of the inhomogeneous Lorentz group, *ISL(2)*, and the internal symmetry group, which we assume here to be *SU(n)*. The linear operator $O^{(1)}(\eta^i)$ is invariant under $IG_4^1 \otimes SU(n)$ transformations of $\eta^i$, $V(\eta^i, \eta^{i'})$ is invariant under simultaneous and identical $ISL(2) \otimes SU(n)$ transformations of $\eta^i$ and $\eta^{i'}$, and $O$ is therefore invariant under simultaneous and identical $ISL(2) \otimes SU(n)$ transformations of all the variables; that is, there is a global $ISL(2) \otimes SU(n)$ invariance.

The linearity of Eq. (1) plus assumption **(2)** guarantee that solutions of the equation can be labeled by the particle-like group representation labels of mass, energy, momentum, spin (all from *ISL*(2)), and charge (from *SU*(*n*)). It also guarantees that when there are several "particles" (particle-like states) present, the usual addition and conservation laws will hold.

**(3)** Because $V(\eta^i, \eta^{i'}) = V(\eta^{i'}, \eta^i)$, the linear operator is invariant under permutations of the sets of variables $\eta^i$. It is presumed, in agreement with the properties of fermion states, that all solutions of physical interest belong to the completely antisymmetric, one dimensional representation of the permutation group.

**Comments on the theory**

*No hidden variables.* Although they are not accessible to experiment, the independent variables bear no relation to the "hidden" variables that are sometimes assumed to underlie quantum mechanics [5], [6], [7, ch. 7]. The conjectured hidden variables uniquely determine the outcome of a given experiment, whereas the underlying independent variables we use have nothing to do with which outcome is perceived. They are simply the substratum in which physical reality takes place.

*Kets.* In this underlying theory, the kets $|M, S, p_\mu, s_z, Q\rangle$ of quantum mechanics, denoted by the group theoretic labels, stand for *functions of the underlying independent variables*.

*Interpretation.* Because the underlying theory is not given in terms of states corresponding to a specific set of elementary particles with definite spin, charge and mass, one might ask what kind of physical universe it refers to. Does it describe a universe composed of objectively existing particles, carrying charge and mass and moving through space, for example? We defer the interpretive question until section 5 and concentrate strictly on the *mathematical* structure of the theory in sections 2-4.

*Space-time.* Roughly speaking, there will be one set of independent variables for each particle-like wave function. These variables, which are complex in the example of the appendix, bear no direct relation to our one time and three space dimensions. So how do our familiar space and time arise in the theory? One can use the generators of space and time translations, $P_\mu$, to map out surfaces of equal space and time in the multi-dimensional independent variable space.

Each generator is a first order differential operator, $f_{i,j}(\eta)\partial_{\eta^{i,j}} = \mathbf{f}\cdot\nabla$, and as such, it defines a direction, $\mathbf{f}/|\mathbf{f}|$, at each point in the independent variable space. Suppose the total number of independent variables is $N_T$. The four f's from the four generators define four directions at each point. Pick an arbitrary point in the independent variable space and call it the space-time origin. One can then map out an $N_T - 4$ dimensional surface which includes the origin and which is orthogonal to the four $P_\mu$-defined directions. Every point on that surface in the independent variable space then corresponds to the single point $x = y = z = t = 0$ of our familiar world. The $P_\mu$ can then be used to step to $N_T - 4$ dimensional surfaces where the three space and one time coordinates are constant but different from 0. In this way, we can cover the whole independent variable space with a grid of $N_T - 4$ dimensional surfaces, each corresponding to a definite value of our space and time coordinates.

*The equations of motion.* From the above construction, where $\Psi$ is defined over the whole independent variable space, we see that $\Psi$ corresponds, not to the state at a particular time, but to the state for all time. The time evolution of the solution can in principle be obtained by using the time translation operator $P_0$ to define surfaces of constant *t* and then examining how the solution $\Psi$ changes from one surface to another.

In practice, however, the time evolution is obtained in a different, representation-based way. First we note that if a solution corresponds to a spin ½ representation that includes both a particle and its antiparticle, the basis vectors can always be chosen so the representatives of the solution obey the Dirac equation. That is, the Dirac equation follows from group representation theory [8, pg 225]. Then the equation of motion for the representation of a fermion, instead of coming directly from Eq. (1), comes from the representation-theory-based free particle Dirac equation. To include interactions, the free particle equation is modified in the way specified by gauge theory. The same holds for vector bosons and the Proca equation [8].

## 3. Representation Theory

It is useful to review the basic elements of representation theory. There are three aspects to it—representations of groups, representations of functions and operators, and the relation of representation theory to the invariance group of a linear operator. As an illustration, consider the equation

(2)
$$\left(\frac{\partial}{\partial u_1}\frac{\partial}{\partial \bar{u}_1} + \frac{\partial}{\partial u_2}\frac{\partial}{\partial \bar{u}_2}\right)\psi = 0$$

where $u_1, u_2$ are complex variables, and the bar denotes complex conjugation. This equation is invariant under the set of all unitary transformations

(3)
$$\begin{bmatrix} u'_1 \\ u'_2 \end{bmatrix} = \begin{bmatrix} a_{11} & a_{12} \\ a_{21} & a_{22} \end{bmatrix} \begin{bmatrix} u_1 \\ u_2 \end{bmatrix} \text{ or } u' = Au, A^* = A^{-1}$$

That is, $\partial_{u'_1}\partial_{\bar{u}'_1} + \partial_{u'_2}\partial_{\bar{u}'_2} = \partial_{u_1}\partial_{\bar{u}_1} + \partial_{u_2}\partial_{\bar{u}_2}$. If we follow one unitary transformation, *A*, by a second, *B*, the result, *C=BA*, is also a unitary transformation, so the set of all 2x2 unitary transformations forms a group, *U*(2).

There are an infinite number of solutions of Eq. (2). These can be grouped together according to which solutions are rotated into each other by a unitary transformation. For example, the three solutions

(4)
$$|1\rangle = u_1^2, |2\rangle = u_1 u_2, |3\rangle = u_2^2$$

are rotated into each other according to

(5)
$$|i\rangle' = |j\rangle R(A)_{ji}$$

where the *R*'s are quadratic functions of the *a*'s ($R_{11} = a_{11}^2$ and so on). In this way, we can associate a 3x3 matrix *R(A)* with every 2x2 matrix *A*, with *R(A)* being the representative of *A*. The representatives obey the same multiplication rules as the original matrices, $C = BA \Rightarrow R(C) = R(B)R(A)$, so the group structure is preserved. The three vectors of Eq. (4) thus form a basis for a *three dimensional representation* of *U*(2).

> More generally, it is the grouping together of sets of solutions of a linear equation that gives group representation theory a good deal of its considerable power in quantum mechanics. The set of all basis vectors for a given representation form a complete set for the problem at hand. In fact, it turns out that the labels on basis vectors—mass, energy, momentum, spin, charge—for a particular representation of the physically relevant group constitute all the particle-like properties of matter we are so familiar with.

In group representation theory, the vectors are labeled by the values of the group invariants and by the eigenvalues of one or more of the generators (generators generate transformations near the identity). In our case, if we ignore the overall phase and look at the generators of *SU*(2), they are

$$(6) \quad \begin{aligned} L_x &= \frac{1}{2}\left(u_2\frac{\partial}{\partial u_1} + u_2\frac{\partial}{\partial u_1}\right) + h.a. \\ L_y &= \frac{-i}{2}\left(u_2\frac{\partial}{\partial u_1} - u_2\frac{\partial}{\partial u_1}\right) + h.a. \\ L_z &= \frac{1}{2}\left(u_1\frac{\partial}{\partial u_1} - u_2\frac{\partial}{\partial u_2}\right) + h.a. \end{aligned}$$

and we use the eigenvalues of $L_z$. The single invariant for $SU(2)$ is $L^2 = L_x^2 + L_y^2 + L_z^2$, which has a value of 2 in the case of the functions of Eq. (4). So the proper group representational labeling of the kets for the functional basis of Eq. (4) is

(7) $\quad u_1^2 = |2,1\rangle,\; u_1 u_2 = |2,0\rangle,\; u_2^2 = |2,-1\rangle$

In addition to having representations of group operations, one can also have representations of more general operators. If we assume the basis vectors $|l\rangle$ are a complete set of states with respect to the operator $O$, where the ket represents or stands for a function of the variables in the problem and the $l$'s are the labels on the functions, then $O|l\rangle = |l'\rangle O_{l',l}$ and $O_{l',l}$ is the (matrix) representative of the operator $O$. One also has representations of functions. If a function is expanded in terms of the complete set $|l\rangle$, $f = \sum \varphi_l |l\rangle$, then $\varphi_l$ is the (vector) representative of the function. The equation $O\Psi = O(\sum \varphi_l |l\rangle) = 0$ becomes, in the representation space, $O_{l,l'}\varphi_{l'} = 0$.

**Representation Theory in Quantum Mechanics.**
The primary invariance group of interest in relativistic quantum mechanics is the inhomogeneous Lorentz group, $IG_4^1$ (or $ISL(2)$). It consists of all "rotations" that leave the form $x^2 + y^2 + z^2 - t^2$ invariant, plus translations of $x, y, z$ and $t$. The labels on representations are mass, $M$, and spin, $S$. The labels on basis vectors within the $(M,S)$ representation are energy, $E$, momentum, $p$ and the z component of spin, $s_z$. There is also an internal symmetry group in quantum mechanics, where "internal" means the group operations have nothing to do with space and time. Labels on vectors for representations of the internal symmetry group (see [9] for a good exposition) include the charges Q—weak, electromagnetic and strong. A state $|\Psi\rangle$ in quantum mechanics can be expanded in terms of functions labeled by $M, S, E, p, s_z$ and $Q$,

(8) $\quad |\Psi\rangle = \sum_l \psi_l |l\rangle = \int \sum \psi(M, S, p_\mu, s_z, Q) | M, S, p_\mu, s_z, Q\rangle$

with the vector representative $\psi(M,S,p_\mu,s_z,Q)$ of the state being the *wave function* in the momentum representation, and the ket $|M,S,p_\mu,s_z,Q\rangle$ representing, in our scheme, a function of the independent variables.

In general, the labels on the kets correspond to the observable characteristics of the state. The independent variables are never observed.

There is one more twist to representation theory. Suppose the linear operator is invariant under permutations and that one wishes to look only at totally antisymmetric representations, as in assumption **(3)** of section 2. Then all states can be expressed as sums of products of anticommuting creation operators acting on the vacuum state and all operators can be expressed as sums of products of creation and annihilation operators, with the creation and annihilation operators denoted by group theoretic labels. For example, the state of Eq. (8) becomes (dropping the *M,S*)

$$(9) \quad |\Psi\rangle = \int \sum \psi(p_\mu,s_z,Q)|p_\mu,s_z,Q\rangle = \int \sum \psi(p_\mu,s_z,Q)a^*(p_\mu,s_z,Q)|0\rangle$$

with $a^*$ being a creation operator and $|0\rangle$ being the vacuum state. Thus field operators are not the "basic mathematical elements" of the underlying theory, as they are in current quantum field theory. Instead, they are a consequence of representation theory applied to the theory outlined in section 2.

## 4. Basis Vectors in the Underlying Theory

We cannot directly convert the underlying, pre-representational theory to the conventional representational form of quantum mechanics because we don't know the specific form of Eq. (1). Nevertheless we can still gain valuable information by using general principles. We envision the conversion from independent variable theory to the current, conventional form of quantum mechanics being done in two steps. The first step is to assume there is a spin ½ basis. And the second is to assume the conventional vacuum, fermion and boson states can be expressed in terms of the spin ½ basis. Note that Nambu and Jona-Lasinio [10] have proposed a theory based on fermion states alone, but their treatment of the vacuum and boson states is different from ours.

### A. The Spin ½ Basis.

For the first step, we assume the basis vectors consists solely of spin ½ states. (Note: This assumption may need to be modified later, but it seemed the most logical way to proceed at this point.) These might, for example, be solutions to the single-bare-particle eigenvector equation

$$(10) \quad \begin{aligned} O^{[1]}(\eta)\psi(\eta:M,E,p,s_z,Q,m) &= \lambda\psi(\eta:M,E,p,s_z,Q,m), \\ s_z &= \pm 1/2, \quad m = 1,\ldots,n \end{aligned}$$

It is assumed that these basis vectors belong to the *n* representation of *SU*(*n*), the set of all *nxn* unitary matrices with determinant 1, with *m* denoting the internal symmetry label. One can Fourier transform to obtain states labeled by position. Solutions will then be sums of products of these states.

**Antisymmetry.**

In accord with assumption **(3)** of section 2, we assume that all states of physical interest are completely antisymmetrized in the sets of independent variables. The effect of this is that one can convert the whole problem into a form where states and operators are expressed in terms of anticommuting spin ½ creation and annihilation operators.

**Renormalization.**

The interaction term in the spin ½ basis will be of the four-fermion form. One might object that such theories are not in general renormalizable [9, pg. 241]—that is, an expansion in terms of Feynman diagrams gives non-cancelable infinities—and so these theories are suspect. However, the use of fermion states alone as a basis is only an intermediate stage in this theory. The final basis will be in terms of vector boson states and the vacuum, as well as fermion states. It is presumably only in this final basis that a Feynman diagram expansion is appropriate. So the renormalization problems encountered when attempting to use a Feynman diagram expansion at an intermediate stage, where the expansion does not seem appropriate, are not, I think, sufficient reason to negate the insights offered by this approach.

## B. The Vacuum State and Vector Bosons.

The second step in converting from an independent variable theory to the conventional form of the theory is to express the usual states of quantum mechanics—the vacuum, vector boson states and so on—in terms of the first, spin ½, basis. The form for the vacuum state, $\Psi_0$, is chosen with an eye to providing an explanation for why a gauge transformation produces a vector boson field proportional to the derivative of the transformation [2]. We will consider here only theories with an unbroken symmetry, although the reasoning can be extended to theories where the vacuum breaks the symmetry of the internal symmetry group [11].

**Gauge Transformations.**

A gauge transformation is a space-time dependent transformation from the internal symmetry group, here assumed to be *SU*(*n*). An infinitesimal gauge transformation can be written as

(11) $\quad G = I + i\varepsilon \int dx \theta^\alpha(x) \tilde{\psi}(x) \tau^\alpha \psi(x)$

with a sum over $\alpha$, where $\tau^\alpha$ is the $\alpha$ th generator of *SU*(*n*), $\psi$ is a column vector of length 4*n* (4 from Dirac, *n* from *SU*(*n*)), and $\theta^\alpha(x)$ tells how the transformation varies in space and time. Suppose we apply this to a state containing a single fermion state superimposed on the vacuum.

$$G\psi * \Psi_0 = G\psi * G^{-1} G \Psi_{vac}$$
(12)
$$= \psi^{*'}(I + i\varepsilon \int dx \theta^\alpha(x) \tilde{\psi}(x) \tau^\alpha \psi(x)) \Psi_{vac}$$
$$= \psi^{*'}(I + i\varepsilon q \int dx \frac{\partial \theta^\alpha}{\partial x_\mu} A_\mu^\alpha(x)) \Psi_{vac}$$

where $\psi^{*'}$ is the rotated fermion state and $A_\mu^\alpha(x)$ is operator for the vector boson field at $x$, with $q$ being the charge. The last line comes from gauge theory, which says that, if we are to match current theory, then a gauge transformation produces a vector boson field proportional to $\partial \theta / \partial x_\mu$. Thus we must have

(13) $$\int dx \theta^\alpha(x) \tilde{\psi}(x) \tau^\alpha \psi(x) \Psi_0 = q \int dx \frac{\partial \theta^\alpha}{\partial x_\mu} A_\mu^\alpha(x) \Psi_0$$

How can we achieve this? The only way is to have the vacuum $\Psi_0$ *not* be invariant under gauge transformations. Instead, a slowly varying gauge transformation must alter the vacuum in such a way as to produce a vector boson field proportional to $\partial \theta^\alpha / \partial x_\mu$.

**Molecular Structure of the Vacuum.**
There is a way to do this. We can construct the vacuum out of "molecules" that are invariant under global internal symmetry transformations from $SU(n)$ but not under local gauge transformations.

$$\Psi_0 = \prod_r \Psi_{cm,r}$$
(14)
$$\Psi_{cm,r} = \int dx_{m,j} F(x_{cm,r}) \prod_{m=1}^{n} \prod_{j=1}^{2} f(x_{m,j} - x_{cm,r}) \tilde{\psi}_m(x_{m,1}) \psi_m(x_{m,2})$$

where $\Psi_{cm,r}$ represents a molecule with center of mass at position $r$, $F$ represents the motion of the center of mass, and $f$ represents the binding of each spin ½ wave function to the molecule (with a force provided by the interaction term of Eq. (1)). The $\psi_m$'s are antisymmetrized four-component spin ½ operators that obeys the Dirac equation. $\tilde{\psi} = \psi * \gamma_0$ consists of a creation operator for particles and an annihilation operator for antiparticles while $\psi$ consists of an annihilation operator for particles and a creation operator for antiparticles. Only the particle and charge conjugate particle *creation* operators survive in Eq. (14). The $\tilde{\psi}_m \psi_m$, and therefore $\Psi_{cm,r}$, are Lorentz spin invariants.

**Vector Bosons.**

To see what happens when the gauge transformation of Eq. (12) is applied to the vacuum of Eq. (14), we expand $\theta(x)$ about $x_{cm,r}$, considered here to be a constant.

(15) $\quad \theta(x) \cong \theta(x_{cm,r}) + (x_\mu - x_{cm,r,\mu})\partial\theta/\partial x_\mu$

where $\theta(x)$ is assumed to vary slowly on the scale of the binding function *f*. The first term, which effectively does not vary over the molecule, gives 0 because of the global invariance of the molecule. The second term, however, will not give 0; instead it will lead to a vector (from the $x_\mu - x_{cm,r,\mu}$) boson (from the $\tilde{\psi}(x)\tau^\alpha\psi(x)$) field. That is, the vector boson fields come from the change induced in the vacuum state by the location-dependent *SU*(*n*) gauge transformation. And the induced boson field is proportional to $\partial\theta/\partial x_\mu$. It also obeys symmetric statistics because it is bilinear in the antisymmetrized fermion creation and annihilation operators.

This construct gives us an understanding of several aspects of elementary particle gauge theory: We see why a gauge transformation produces a vector boson field; we see why the field is proportional to $\partial\theta/\partial x_\mu$; we see why vector bosons obey symmetric statistics; we see why there is one gauge field for each generator of the invariance group; and we see why the gauge fields have the transformation properties of the generators of the invariance group. (See appendix B for an attempt at a derivation of gauge invariance.)

**The Physical Spin ½ States.**

We have sketched the construction of the vacuum and the vector boson states in terms of the more basic *SU*(*n*) spin ½ states. What about the physical spin ½ states, the quarks, electrons and neutrinos? We conjecture that they will consist of modified vacuum molecules with an odd number of basic spin ½ states. We also conjecture that the different generations would come from different excitation states of the molecules corresponding to the fermion states.

As an example of an *SU*(*n*)-based model, we could use *SU*(6). The first three basic states, $|q,1\rangle, |q,2\rangle, |q,3\rangle$, are quark-like, but with no weak interaction. The next two, $|\nu\rangle, |e\rangle$, are like the neutrino and electron leptons, and the last, $|x\rangle$ has no strong or electro-weak charge. The *SU*(6) structure of the physical quarks would be triplets composed of one basic quark, one lepton or anti-lepton, and one $|x\rangle$. But this is not meant to imply that quarks are composites of three basic particles (because this gives trouble with particle-antiparticle annihilation).

# 5. Interpretation of the Underlying Theory.

At its most basic level, as is illustrated in the complex variable example of appendix A, there is no space-time and there are no particles in this conjectured underlying theory. Further, one has the usual interpretive difficulties that bedevil the conventional, linear, many-versions-of-reality quantum mechanics. So how are we to understand this theory? How does it mesh with our perceptions of the world?

First, space-time. The construction of section 2 shows how our familiar three space and one time dimension are introduced into the theory. In a reversal of the usual procedure, where Lorentz transformations are defined on a pre-existing space-time, the translation generators of *ISL*(2) can be used to superimpose a space-time structure on the underlying independent-variable theory. It is this superimposed structure that we, in effect, perceive.

Second, the particle-like nature of physical existence. There are no particles in this theory; all that "exists" are the state vectors, which are functions of the independent variables. But because of the Lorentz invariance and the internal symmetry group, these state vectors will have particle-like properties—mass, energy, momentum, spin, and charge. These effectively give the state vectors the properties of particles, and so we perceive the physical world as particle-like.

Third, there is the usual difficult problem of the interpretation of quantum mechanics, with its many versions of reality. I would define this problem in the following way: Suppose we start out with just the linear, relativistic mathematics of quantum mechanics—no objectively existing particles, no collapse, no "sentient beings"—and ask whether this bare "many-worlds" structure is sufficient for a proper description of physical reality. We find that it is not. If the probability law is to hold, then *something* must be added to this simple scheme, with that "something" *singling out* one version of the state vector for perception [12]. There are, I believe, only three ways to single out a version.

The first is to use "hidden variables" whose specific values on a given run of an experiment single out just one version. There is no experimental evidence for hidden variables, but some theoretical work has been done. In Bohm's well-known example [5], [6], the hidden variables are the positions and velocities of objectively existing particles. These particles are constrained by their equations of motion to ride along on just one version of the state vector, in such a way that the probability law is satisfied. But there are problems with this approach [13] (see also [14] for no particles by implication; and it would seem that all collapse schemes also implicitly assume no particles). The first, specific to the Bohm model but difficult to remedy in any model, is that the *mathematics* does not specify how many particles are to be associated with a "single-particle" wave function; it must be *assumed* that there is only one particle per single-particle wave function. The second problem is that it is not clear that singling out one version in *any* hidden variable model solves the problem of perception of just one branch. For there will be valid versions of the wave function of the observer's brain on all branches, and there is no way, within the models, to show why all the non-singled-out versions are not "aware."

So it appears to be extremely difficult, if not impossible, to construct an acceptable hidden variable theory which meshes with quantum mechanics. And it is just as well for our underlying theory that hidden variables cannot be shown to work because they are not present in the mathematics of section 2. (The values of the conjectured hidden variables determine the perceived outcome, but that is not true of the independent variables; they don't take on a specific value on a given run.)

The second way to single out a version is by having the wave function/state vector collapse down to just one branch. Ghirardi, Rimini, and Weber [15] and Pearle [16]-[19] in particular have proposed a most elegant scheme. There are, however, problems with their scheme [20]. There is no experimental evidence for collapse. There must be instantaneous coordination between distant parts of the wave function, and even between parts in different universes. In addition, the underlying quantum mechanics must be non-linear [20, appendix A]. Thus again, it is just as well for our scheme that collapse (as far as we know) does not solve the interpretive problem, for our proposed theory could not accommodate collapse because it is linear.

So the only apparent possibility left is to suppose that the mathematics of linear, relativistic, non-collapse, no-hidden-variable quantum mechanics constitutes the entire mathematics that is necessary to describe physicality. This is the premise of many-worlds interpretations [21]-[26]. These interpretations, however, do not single out one version of the observer as *the* perceiving version. But one can show that the probability law requires that one version is indeed singled out [12]. Thus many-worlds interpretations are inadequate. To remedy this, one must make the further assumption [12] that each observer has a "perceiving aspect," outside the mathematics of quantum mechanics, that looks into physical reality (consisting of the state vectors) and perceives just one branch of the state vector. Pictorially, each branch corresponds to a different location in the independent variable space, and "awareness" can be in only one of those locations.

It is no doubt distasteful to most physicists to suppose that an aspect ("awareness") outside the mathematics is necessary to bring quantum mechanics into agreement with our perceptions. But the apparent need for this assumption is not specific to our underlying theory model; unless one finds evidence for hidden variables (particles) or collapse, it is also needed for the understanding of the conventional formulation of quantum mechanics. (I am not the first physicist to propose this; the Nobel laureate E. P. Wigner long ago [27] proposed a similar "sentient being" scheme, only he used collapse, and that is not necessary.)

## 6. Summary and Conclusions.

The current theory of quantum mechanics depends heavily on concepts from group representation theory:

- •The particle-like properties of mass, energy, momentum, spin and its z-component are all labels on basis vectors for representations of the inhomogeneous Lorentz group. And charges are labels on basis vectors for representations of the internal symmetry group.
- The experimentally verified addition laws for all these particle-like properties follow from group representation theory, and the conservation laws follow from invariance assumptions.
- With a suitable choice of basis, the spin ½ Dirac and spin 1 Proca free-particle equations follow from representation theory.
- The antisymmetry of fermions is a property associated with representations of the permutation group.
- The vector bosons which mediate interactions transform as the generators of the internal symmetry group.
- The mathematics of the internal symmetry group figures strongly in the highly successful Standard Model [9], [28], [29] of elementary particles.
- Gauge theory is closely connected to the internal symmetry group.

All these representation-theory related properties strongly suggest that the current form of quantum mechanics is a representation of an underlying, pre-representational form of the theory. The conjecture here is that there is an underlying, pre-representational linear equation, perhaps a partial differential equation, in some set of currently unknown independent variables (which are not "hidden" variables), with the linear operator being invariant under a suitable group of transformations of the variables. Such an underlying theory would presumably not affect any of the successes of current quantum mechanics, and it would explain the ubiquitous group representational structure.

We have given an example with an $SU(n)$ internal symmetry group in which the independent variables are complex. Space-time variables $x_\mu$ are introduced by use of the generator $P_\mu$ of translations.

It is proposed that the conversion to the current, representational form of quantum mechanics is carried out in two steps. In the first step, spin ½ basis vectors are used. All states, including the vacuum and vector bosons, are then constructed from sums of antisymmetrized products of these states.

Gauge transformations are presumed to correspond to a change of basis. Infinitesimal gauge transformations disturb the vacuum, producing vector gauge bosons which obey commutation (rather than anticommutation) rules. From this construction, one gains an understanding of why a gauge transformation produces a gauge field, why the field is proportional to $\partial\theta/\partial x_\mu$, why there is one gauge field for each generator of the invariance group, and why the gauge fields have the transformation properties of the generators of the invariance group. A tentative "derivation" of gauge invariance is given in appendix B.

It is conjectured in appendix C that general relativity could be fit into this scheme by supposing that macroscopic concentrations of "matter" macroscopically alter the structure of the vacuum, and this altered structure leads to curved space-time.

Finally, there is the problem of the interpretation of a physical theory in which there are no particles, there is no space-time in its initial formulation (these properties appear in the *solutions*), and there are the usual multiple versions of reality. Hidden variable and collapse interpretations are not suitable for the interpretation of an underlying theory, so we are forced to an interpretation that lies outside the mathematics of the theory. (But with no evidence for particles or collapse, we are led to the same conclusion in the conventional formulation of quantum mechanics.)

This paper gives only a sketch of an underlying theory, with the major part of the work still to be done. In particular, the correct underlying equation needs to be found, and the conversion of the pre-representational equation to representational form needs to be carefully thought out.

# Appendix A.
# An Example of an Underlying Theory.

### 1. The Single-Particle Equation.

It is useful to have an example of an underlying theory so that one can see the possibilities. We will give one here [3], in which the independent variables are complex variables (unlike space and time which are real variables). The reason complex variables are preferred is that it seems easier to include complex internal symmetry groups.

We start with the equation for a single particle-like state, having mass, spin, charge and so on. It is a linear, second order partial differential equation that has the same general structure as the harmonic oscillator problem in quantum mechanics, only in this case the harmonic oscillator is relativistic. This form was chosen because one can find explicit solutions. The equation is

(A-1) $O^{[1]}\Psi = 0$

(A-2) $O^{[1]} = -\left( \dfrac{\partial}{\partial u_{1i}} \dfrac{\partial}{\partial v_{2i}} - \dfrac{\partial}{\partial v_{1i}} \dfrac{\partial}{\partial u_{2i}} + \dfrac{\partial}{\partial \bar{u}_{1i}} \dfrac{\partial}{\partial \bar{v}_{2i}} - \dfrac{\partial}{\partial \bar{v}_{1i}} \dfrac{\partial}{\partial \bar{u}_{2i}} \right)$
$\qquad\quad + \left( u_{1i}v_{2i} - v_{1i}u_{2i} + \bar{u}_{1i}\bar{v}_{2i} - \bar{v}_{1i}\bar{u}_{2i} \right)$

There is a sum on *i* from 1 to *n* (with *n* on the order of 6), and the bar denotes complex conjugate.

### 2. Symmetry Operations.
**The Lorentz Group**

There are 2 real variables for each $u_{1i}, v_{1i}, u_{2i}, v_{2i}$ for a total of 8 per *i*. Thus there are a total of $8n$ real variables. If we switch to variables in which $O$ is diagonal, it will be a quadratic form with half the coefficients +1 and half the coefficients –1.

Thus the $O$ of Eq. (A-2) has a symmetry group with $16n^2 + 4n - 1$ generators. For now, we are only interested in those generators that correspond to the inhomogeneous Lorentz group. They must satisfy the commutation relations

(A-3a) $\quad [J_i, J_j] = i\varepsilon_{ijk} J_k, \quad [J_i, K_j] = i\varepsilon_{ijk} K_k, \quad [K_i, K_j] = -i\varepsilon_{ijk} J_k$

(A-3b) $\quad [J_i, P_j] = i\varepsilon_{ijk} P_k, \quad [K_i, P_j] = i\delta_{ij} P_0, \quad [J_i, P_0] = 0, \quad [K_i, P_0] = iP_i$

(A-3c) $\quad [P_i, P_j] = 0, \quad [P_i, P_0] = 0$

The generators of the group $SL(2)$, the set of all 2x2 complex matrices with determinant 1, homomorphic to the homogeneous Lorentz group, are

(A-4a)
$$J_1 = \frac{1}{2}\left( u_{bi} \frac{\partial}{\partial v_{bi}} + v_{bi} \frac{\partial}{\partial u_{bi}} - \bar{u}_{bi} \frac{\partial}{\partial \bar{v}_{bi}} - \bar{v}_{bi} \frac{\partial}{\partial \bar{u}_{bi}} \right)$$

$$J_2 = \frac{i}{2}\left( u_{bi} \frac{\partial}{\partial v_{bi}} + v_{bi} \frac{\partial}{\partial u_{bi}} - \bar{u}_{bi} \frac{\partial}{\partial \bar{v}_{bi}} - \bar{v}_{bi} \frac{\partial}{\partial \bar{u}_{bi}} \right)$$

$$J_3 = \frac{1}{2}\left( u_{bi} \frac{\partial}{\partial u_{bi}} - v_{bi} \frac{\partial}{\partial v_{bi}} - \bar{u}_{bi} \frac{\partial}{\partial \bar{u}_{bi}} + \bar{v}_{bi} \frac{\partial}{\partial \bar{v}_{bi}} \right)$$

$$K_1 = \frac{i}{2}\left( u_{bi} \frac{\partial}{\partial v_{bi}} + v_{bi} \frac{\partial}{\partial u_{bi}} + \bar{u}_{bi} \frac{\partial}{\partial \bar{v}_{bi}} + \bar{v}_{bi} \frac{\partial}{\partial \bar{u}_{bi}} \right)$$

$$K_2 = -\frac{1}{2}\left( -u_{bi} \frac{\partial}{\partial v_{bi}} + v_{bi} \frac{\partial}{\partial u_{bi}} + \bar{u}_{bi} \frac{\partial}{\partial \bar{v}_{bi}} - \bar{v}_{bi} \frac{\partial}{\partial \bar{u}_{bi}} \right)$$

$$K_3 = \frac{i}{2}\left( u_{bi} \frac{\partial}{\partial u_{bi}} - v_{bi} \frac{\partial}{\partial v_{bi}} + \bar{u}_{bi} \frac{\partial}{\partial \bar{u}_{bi}} - \bar{v}_{bi} \frac{\partial}{\partial \bar{v}_{bi}} \right)$$

where $b$ is summed from 1 to 2 and $i$ from 1 to $n$. And the generators of translations are

(A-4b)
$$P_0 = u_{1i} \frac{\partial}{\partial \bar{v}_{2i}} - v_{1i} \frac{\partial}{\partial \bar{u}_{2i}} - \bar{u}_{1i} \frac{\partial}{\partial v_{2i}} + \bar{v}_{1i} \frac{\partial}{\partial u_{2i}}$$

$$P_1 = -u_{1i} \frac{\partial}{\partial \bar{u}_{2i}} + v_{1i} \frac{\partial}{\partial \bar{v}_{2i}} + \bar{u}_{1i} \frac{\partial}{\partial u_{2i}} - \bar{v}_{1i} \frac{\partial}{\partial v_{2i}}$$

$$P_2 = i\left( u_{1i} \frac{\partial}{\partial \bar{u}_{2i}} + v_{1i} \frac{\partial}{\partial \bar{v}_{2i}} + \bar{u}_{1i} \frac{\partial}{\partial u_{2i}} + \bar{v}_{1i} \frac{\partial}{\partial v_{2i}} \right)$$

$$P_3 = u_{1i} \frac{\partial}{\partial \bar{v}_{2i}} + v_{1i} \frac{\partial}{\partial \bar{u}_{2i}} - \bar{u}_{1i} \frac{\partial}{\partial v_{2i}} - \bar{v}_{1i} \frac{\partial}{\partial u_{2i}}$$

Note that there is no scale in the momenta. One could multiply by anything and the commutation relations still work. The scale must come from the interaction.

We can also give the Lorentz transformations macroscopically. The homogeneous Lorentz transformation $L(A)$ corresponding to the 2x2 matrix $A$ from $SL(2)$ has the effect (we drop the subscripts here)

$$\text{(A-5a)} \quad L(A)\begin{bmatrix} u \\ v \end{bmatrix} = \begin{bmatrix} a_{11} & a_{12} \\ a_{21} & a_{22} \end{bmatrix}\begin{bmatrix} u \\ v \end{bmatrix} = \begin{bmatrix} a_{11}u + a_{12}v \\ a_{21}u + a_{22}v \end{bmatrix}$$

while the effect of translations is

$$\text{(A-5b)} \quad T(x_\mu)\begin{bmatrix} u_1 \\ v_1 \\ u_2 \\ v_2 \end{bmatrix} = e^{iP_\mu x_\mu}\begin{bmatrix} u_1 \\ v_1 \\ u_2 \\ v_2 \end{bmatrix} = \begin{bmatrix} u_1 \\ v_1 \\ u_2 - ix_0\bar{v}_1 + ix_1\bar{u}_1 - x_2\bar{u}_1 - ix_3\bar{v}_1 \\ v_2 + ix_0\bar{u}_1 - ix_1\bar{v}_1 - x_2\bar{v}_1 - ix_3\bar{u}_1 \end{bmatrix}$$

**Unitary Transformations $SU(n)$.**

It is useful to introduce an $SU(n)$ "internal" symmetry, where $SU(n)$ consists of the set of all $n\times n$ unitary matrices with determinant 1. If the kets $|j\rangle$ transforms as the $n$ representation of $SU(n)$ and the kets $|\bar{j}\rangle$ as the $\bar{n}$ representation, then $|j\rangle|\bar{j}\rangle$ (summed on $j$ from 1 to $n$) is invariant under $SU(n)$. In order to make the generators of the inhomogeneous Lorentz group $SU(n)$ invariants, we suppose the variables with $b=1$ transform as the $n$ representation and the variables with $b=2$ transform as the $\bar{n}$ representation. The $SU(n)$ transformation properties of all the variables and their derivatives are then

$$\text{(A-6)} \quad \begin{aligned} &|j\rangle : u_{1j}, v_{1j}, \partial\bar{u}_{1j}, \partial\bar{v}_{1j}, \bar{u}_{2j}, \bar{v}_{2j}, \partial u_{2j}, \partial v_{2j} \\ &|\bar{j}\rangle : u_{2j}, v_{2j}, \partial\bar{u}_{2j}, \partial\bar{v}_{2j}, \bar{u}_{1j}, \bar{v}_{1j}, \partial u_{1j}, \partial v_{1j} \end{aligned}$$

So we see that the operator of Eq. (A-2) and the generators of the inhomogeneous Lorentz group are invariants under $SU(n)$. Charges will correspond to the elements of the $n-1$ diagonal generators of $SU(n)$

### 3. Particlelike States.

We will not go through the details of the mathematics, but one can show that there are particlelike solutions to Eq. (A-1,2). There are solutions for any nonzero mass and any spin. One of the unusual features is that the normalization,

(A-7) $\langle \psi_{p_\mu}(u,v) | \psi_{p'_\mu}(u,v) \rangle = \delta^4(p_\mu - p'_\mu)$

involves a $\delta^4$ rather than the expected $\delta^3$. The reason for this is that one of the generators of the invariance group obeys $[G, P_\mu] = i\lambda P_\mu$ so that it can be used to scale the mass. That is, if $\psi$ is a solution to the equation with mass $m$, then for any $\beta$, $e^{-i\beta G}\psi$ is also a solution (because $G$ commutes with $O$), and it has mass $\beta\lambda m$.

### 4. An Equation with Interactions.

One can modify this complex-variable model to include interactions. To do this, instead of a single set of $u_{bi}, v_{bi}$, $b=1,2$, $i=1,...,n$, we use $N$ sets, $u_{bi}^{(m)}, v_{bi}^{(m)}$, $m=1,...,N$. Then we set the linear operator equal to

$$O = O^{(1)} + O^{(2)}$$

(A-8) $\quad O^{(1)} = \sum_{m=1}^{N} O^{[1],m}$

$$O^{(2)} = \sum_{m' \neq m} V(u^{(m)}, v^{(m)}; u^{(m')}, v^{(m')}) = \sum_{m' \neq m} V(m, m')$$

$O^{[1],m}$ is the operator of Eq. (A-2), using the $m$th set of $u,v$, and $V(m,m') = V(m',m)$ so that $O$ is invariant under permutations of sets of variables.

The generators of the invariance group are now the sums of the generators for the individual sets of variables (Eqs. (A-4) and (A-5)). To have an invariant theory, $O^{(2)}$ must commute with those generators. In particular, it must commute with the total momentum operators

(A-9) $\quad P_\mu = \sum_{m=1}^{N} P_\mu^{(m)}$

In order to have the interaction transfer momentum from one "single-particle" state to another, and yet conserve total momentum, we must have

(A-10) $P_\mu^{(m)} V(m,m') \neq 0$, $P_\mu^{(m')} V(m,m') \neq 0$, $\left(P_\mu^{(m)} + P_\mu^{(m')}\right) V(m,m') = 0$

We can achieve this in the following way: Let

(A-11) $-iP_2(i,m) = u_{1,i,m} \dfrac{\partial}{\partial \bar{u}_{2,i,m}} + v_{1,i,m} \dfrac{\partial}{\partial \bar{v}_{2,i,m}} + \bar{u}_{1,i,m} \dfrac{\partial}{\partial u_{2,i,m}} + \bar{v}_{1,i,m} \dfrac{\partial}{\partial v_{2,i,m}}$

(A-12)
$$I(1,j,n:2,\bar{j}',n') = u_{1,j,n}v_{2,j',n'} - v_{1,j,n}u_{2,j',n'}$$
$$\bar{I}(1,\bar{j},n:2,j',n') = \bar{u}_{1,j,n}\bar{v}_{2,j',n'} - \bar{v}_{1,j,n}\bar{u}_{2,j',n'}$$

where the bar over the $j$ or $j'$ indicates the $SU(n)$ transformation properties.

Then

(A-13)
$$-iP_2(i,m)I(1,j,n:2,\bar{i},m) = u_{1,j,n}\bar{v}_{1,i,m} - v_{1,j,n}\bar{u}_{1,i,m} = X_2(j,n:\bar{i},m)$$
$$-iP_2(i,m)\bar{I}(1,\bar{j},n:2,i,m) = \bar{u}_{1,j,n}v_{1,i,m} - \bar{v}_{1,j,n}u_{1,i,m} = -X_2(i,m:\bar{j},n)$$
$$-iP_2(j,n)\bar{I}(1,\bar{i},m:2,j,n) = \bar{u}_{1,i,m}v_{1,j,n} - \bar{v}_{1,i,m}u_{1,j,n} = -X_2(j,n:\bar{i},m)$$

So if we set

(A-14) $\quad J(i,\bar{j}:m,n) = I(j,n':\bar{i},m) + \bar{I}(\bar{i},m:j,n)$

then

(A-15) $\quad \left[\sum_{k,m'} P_\mu(k,m')J(i,\bar{j}:m,n)\right] = \left(P_\mu(i,m) + P_\mu(j,n)\right)J(\bar{i},j:m,n) = 0$

So any function of the $J$'s will have net zero momentum. Thus we can construct $V(m,m')$ from arbitrary functions of these invariants, provided it is invariant under exchanges of $m$ and $m'$. For example, we could use the quartic form

(A-16) $\quad V(m,m') = \sum_{i,j} J(i,\bar{j}:m,m')J(j,\bar{i}:m',m) = V(m',m)$

to obtain an $SU(n)$ invariant (because $\sum|i\rangle|\bar{i}\rangle, \sum|j\rangle|\bar{j}\rangle$ are $SU(n)$ invariants).

Since the interaction term is invariant under exchanges of $m$ and $m'$ variables, the operator $O$ is invariant under the permutation group. Thus, in accord with the antisymmetry of fermions, we can restrict our attention solely to those solutions that are totally antisymmetric under exchange of sets of underlying variables.

**The Variational Principle.**

Quantum field theory is usually cast in terms of a variational principle, $\delta L = 0$. In the current context, it is relevant to note that any linear equation can be recast in terms of a variational problem:

(A-17) $\quad O\Psi(\eta) = 0 \Leftrightarrow /\delta \int d\eta \, \bar{\Psi}O\Psi \,/\delta\bar{\Psi} = 0$.

# Appendix B.
# Gauge Invariance.

## 1. Yang-Mills Gauge Theory.

The pivotal paper on gauge transformations was that of Yang and Mills [2] in which the internal symmetry group was taken to be *SU(2)* (isotopic spin). In that paper, they say "We define *isotopic gauge* as an arbitrary way of choosing the orientation of the isotopic spin axes at all space-time points…. We then propose that all physical processes be invariant under an isotopic gauge transformation $\psi \to \psi', \psi' = S^{-1}\psi$ where *S* represents a space-time dependent isotopic spin rotation." We can infer more detailed, specific assumptions from this:

- At each point in space, there is a set of *n* axes related to the internal symmetry group *SU(n)*. There is no real understanding, in my opinion, of what the axes refer to in the conventional formulation of quantum mechanics.
- The internal symmetry states of all particle states—all fermions, all bosons—are measured with respect to the *same* set of axes.
- The equations of quantum mechanics are assumed to be invariant under arbitrary rotations of the axes at each point in space. This does not follow from any other mathematically established principle; it is a separate assumption.

The principle of gauge invariance almost completely determines the form of the interactions in elementary particle physics, from the replacement of the usual derivative with the covariant derivative to the zero mass of vector bosons associated with non-broken symmetries. We know from its far-reaching, verified consequences that the results of the Yang-Mills proposal on gauge invariance are correct (when applied to the appropriate internal symmetry group). However, since a gauge transformation is not part of the invariance group, I do not see any group theoretic rationale for the basic assumption of gauge theory—that the forms of the equations remain invariant under gauge transformations. It is certainly a convincing plausibility argument, but it is not a proof of invariance of the form of the equations under gauge transformations.

## 2. Gauge Invariance When Gauge Transformations Are Generated by a Change of Basis.

We will give an argument for gauge invariance based on the ideas of section 4, where gauge transformations are a change of (fermion) basis. We leave it to the reader to decide whether it is actually a proof of invariance.

A physical state, $\Psi$, correspond to a solution of Eq. (1). If the basis states are $\Phi_l$, then

(B-1) $\quad \Psi(\eta) = \int dl\, \psi_l \Phi_l(\eta)$

where the $l$'s are labels on states, and the $\psi_l$ are the representatives of $\Psi$ in that particular basis. It is the $\psi_l$ that are the objects in the current theory, so the equations of motion will be equations (in terms of the labels $l$) for the $\psi_l$. If we use a different set of basis function, $\Phi'_l$, perhaps using a gauge transformation to generate the new basis, then the state will be the same, but the representative $\psi'_l$ will be different.

(B-2) $\quad \Psi = \int dl\, \psi'_l \Phi'_l$

The question is whether the equations of motion for $\psi'_l$ will have the same form as the equations for $\psi_l$.

Because $\psi_l$ and $\psi'_l$ refer to the same solution of the underlying equation, the solutions to the *representative* equations, one for $\psi_l$ and one for $\psi'_l$, must refer to *the same* physical state. Thus if the two representative equations are different, the difference cannot lead to any *physical* difference in the sets of solutions of the two equations (because $\Psi$ completely determines the physical content). So the best we can do is to say that the differences in form (such as the addition of a gauge-induced field) of the equations for $\psi_l$ and $\psi'_l$ cannot have any physical significance; that is, the differences in form cannot lead to states with different physical characteristics. Perhaps that is sufficient.

## Appendix C.
## Gravity.

It would be satisfying if this underlying theory also included gravity. Is there any way this could come about?

First, one can make a general argument that gravity should be directly connected with quantum mechanics. Gravity is, in a sense, a theory about mass. But mass is a child of (special relativistic) quantum mechanics because mass follows, via group representation theory, from the Lorentz invariance and linearity of quantum mechanics. Also the Higgs boson is presumed to give mass to particles. So we would be a little surprised if there were not a close connection between gravity and "conventional" quantum mechanics.

How might this be implemented? We know that particles—quarks, electrons and so on—polarize the vacuum; that is, they affect the microscopic structure of the vacuum. And so it is conceivable that a macroscopic concentration of particles might slightly change the macroscopic structure of the vacuum. For example, it might slightly change the density of particlelike functions that make up the vacuum. The smallness in the change of the structure of the vacuum could account for the smallness of the gravitational constant.

One need not solve the whole vacuum problem to set this up. One probably only needs certain properties of the "single-particle" density matrix, perhaps the local energy-momentum density. That is, the proposal is that gravity is a macroscopic theory of the vacuum in the presence of concentrations of matter. Thus, except for the fact that the $P_\mu$ are used to define the $x_\mu$, this proposal is somewhat independent of the material in the rest of the paper.

**Space-time.**

How would one derive the gravitational equations? I am not certain. But it would presumably have something to do with the way in which a space-time grid is superimposed on the space of independent variables (section 2). In a vacuum, when there are no concentrations of matter, the vacuum state is invariant under the $P_\mu$ and the $x_\mu$ are defined by $P_\mu(\eta) x_\nu(\eta) = \pm \delta_{\mu\nu}$. But when there are concentrations of matter then (presumably) the vacuum state is not invariant under the "local" space-time generators and, very roughly, the $x_\mu$ are defined by something like $P_\mu(\eta)\bigl(x_\nu(\eta)\Psi_{vac}(\eta)\bigr) = \pm \delta_{\mu\nu}\Psi_{vac}(\eta)$. From this plus assumption about the properties of the vacuum, I conjecture that one could derive the general relativistic equations connecting space, time, and densities of matter. (Note that the local $P_\mu$ has two roles; it generates the local $x_\mu$, and its expectation value is proportional to the local energy-momentum density.) This scheme, where space, time and matter are "emergent" rather than "fundamental" properties (that is, they are properties of the *solutions* rather than being built directly into the original equation) may make it easier to conceptually understand how matter could alter space and time.

# References.